\documentstyle[eqsecnum,aps,epsf,prl]{revtex}
\def\reac{$p+\alpha\to p+ \pi^0 +\alpha$}

\begin{document}
\title{Relevant polarization observables for the study of the Roper resonance 
with hadronic probes.}
\author{Egle Tomasi-Gustafsson}
\address{ DAPNIA/SPhN, CEA/Saclay, 91191 Gif-sur-Yvette Cedex, 
France}
\author{Michail P. Rekalo \footnote{ Permanent address:
\it National Science Center KFTI, 310108 Kharkov, Ukraine}
}
\address{Middle East Technical University, 
Physics Department, Ankara 06531, Turkey}
\maketitle
\date{\today}
\begin{abstract}
The purpose of this contribution is to stress the great importance of 
polarization observables for the study of nucleon resonances, in particular in 
the case of deuteron and proton probes. Polarization observables play a unique 
role 
in signing the reaction mechanism and bring new information in the microscopic 
description of baryon resonances. We will calculate the relevant 
polarization observables, for $p+d$- and $p+\alpha$-collisions,
in case of inclusive reactions and for some selected exclusive channels.
\end{abstract}
\section{Introduction}

After the static properties as 
masses, widths and magnetic moment,  a good comprehension of  
the form factors in the ground state as well as in the excited state, is 
required.
In the field of nucleon resonances, in particular, the nature of the nucleonic 
resonance $P_{11}(1440)$, the 
{\it Roper resonance}, with isotopic spin $1/2$ and usual spin $1/2$
(excited nucleon) is still not well known. This resonance was
discovered during the phase shift analysis of data on 
$\pi N$-scattering \cite{Ro64}, it is predicted by most of the quark models, 
but 
never directly observed. 

The measurement of transition nucleon form factors is mostly based on studies 
with 
electromagnetic 
probes where one measures the different polarization observables for 
meson electroproduction processes. However, due to the isospin nonconservation 
in the electromagnetic interaction, a 
photon 
does not have a definite value of isospin.
Therefore it is not possible to separate the isoscalar and 
isovector contributions, using only data about inclusive $\gamma^* p-$and 
$\gamma^* n-$collisions, but, due to the isoscalar selectivity of the 
deuteron probe
(in the framework of a $t-$channel $\omega-$exchange approximation
for the process $\vec d+p\to d+X$), it is possible to determine the isoscalar
amplitude  separately.
In this respect hadronic processes may be considered as the necessary and 
complementary tools to study the isotopic structure of $\gamma^* 
N-$interactions.

The study of $N^*-$structure by hadronic probe has known advantages with respect 
to real or virtual photons: large values of cross sections, advanced technics 
of 
high intensity $\vec d$ and  $\vec p$ beams, polarized targets and polarimeters, 
absence of problem of radiative corrections, natural selection of isoscalar 
$N^*$-excitation, if one chooses isoscalar particles, as $d$ or $\alpha$. On the 
other hand, the reaction mechanism is not well known $a~priori$. 
The $\omega-$exchange seems to be the best mechanism  to describe the 
$N^*$-excitation in $dp-$ or $d\alpha-$collisions: the $\omega NN$ coupling 
constant is large, a spin one exchange 
allows to obtain very specific 
polarization phenomena and energy-independent cross section. In the framework 
of this mechanism, it is possible to predict all observables 
for the reaction $d+p\rightarrow d+X$ in terms of deuteron electromagnetic 
form 
factors and isoscalar form factors of $N\rightarrow N^*$ transitions 
\cite{Re96,ETG99}.

In this contribution we will recall the properties of the $d+p\rightarrow d+X$  
and $\alpha +p\rightarrow \alpha+X$ reactions in the framework of $\omega-$ 
($\sigma-,~\eta$-) exchange and derive the relevant polarization observables.

\section
{The reaction {\boldmath $\lowercase{d}+\lowercase{p}\to \lowercase{d}+X$}}

The 
differential cross section has been measured in the past  at incident 
energy $E_d$=1.6 GeV \cite{Ba73} and more recently at $E_d$=2.3 GeV, at the 
Saturne 
accelerator \cite{lns250}. Polarization observables have been measured at 
Saturne 
\cite{lns250} and Dubna \cite{Az96}. 
The existing data on the 
differential cross section for this reaction show the presence of at 
least two mechanisms in the intermediate energy region ( $2\leq E_{kin}\leq 
10~GeV$). One is the coherent excitation of $d$ with pion production (Fig. 
1a), 
which results in a Deck peak \cite{De64}, in the energy spectrum of the 
scattered deuterons.
The isotopic spin of this peak is $I=1/2$, but the spin ${\cal J}$ and the 
space 
parity $P$ of the Deck peak may not have a unique value. The Deck effect
decreases when the energy of the colliding particles increases, while the role 
of 
a second mechanism, the direct $N^*-$ excitation, must become more important. 
This 
can be described by a $t-$channel exchange by mesonic states (Fig. 2b), with 
$I=0$ 
and with different ${\cal J}^P$: $\sigma,~\eta,~\omega..$.

The tensor analyzing powers, which are strongly negative, show 
a large similarity with the tensor polarization data measured in $ed$ elastic 
scattering \cite{t20}, therefore suggesting the $\omega-$meson being the most 
probable mediator. 

An extended version of a model based on $t-$channel $\omega$-exchange  has been 
published in \cite{Re96,ETG99}.
Let us recall here two main results: the linearity of the cross section 
as a function of a kinematical variable $y$, defined below, and a 
factorization of the tensor analyzing power in a function of the deuteron and 
the $N^*$ electromagnetic form factors.

\subsection{General structure of the differential cross-section}

The differential cross section, corresponding to diagram $1b$ 
(with $\omega$-exchange), can be 
written in the following form:
\begin{equation}
\frac {d^2\sigma}{dtdw^2}=N[y^2+a(t,w^2)],
\end{equation}
where: $y=\displaystyle \frac{p_1\cdot p_2 -k\cdot p_2/2}{mM},~t=(p_1-p_3)^2$ 
and $w$ is the invariant effective mass of $X$, in the process $d+p\to d+X$.
Here $m$ ($M$) is the proton (deuteron) mass. The notation of four-momenta is 
illustrated in Fig. 1b.

The linear $y^2-$dependence of the differential cross section is a direct 
consequence 
of the spin one $\omega-$exchange in the $t-$channel. In a general case the 
inclusive scattering process $p(d,d)X$ is characterized by 3 independent 
variables: $t,w^2$ and $s$, $s=(p_1+p_2)^2$. The dependence of the cross section 
from $t$ and 
$w^2$ has a dynamical character and is described by the corresponding SF's for 
the $dd\omega-$ and $\omega pX-$ vertexes. 

The linearity of the dependence (1) of the double differential cross section
on $y^2$ can be experimentally tested by a measurement of the differential cross 
section, at three different 
energies of the initial deuteron (3 different values of $s$), at fixed values 
of $t$ and $w^2$. This linearity is a direct consequence of the 
$\omega-$exchange mechanism, and such measurement would be an experimental 
test of the validity of the $\omega-$exchange mechanism, equivalent to the 
Rosenbluth fit for electron-hadron scattering.  
One can mention that the Rosenbluth fit allows also to separate the 
contributions of longitudinal and transversal photon polarizations to the 
differential cross section of any process $e^-+A\rightarrow e^-+X$. Similarly, 
the study of the $y^2-$linearity of the double differential cross section for 
processes $d+p\rightarrow d+X$ will allow to separate two different 
structure functions. At the limit $s \gg w^2,~ |t|$, we have:
\begin{equation}
\frac {d^2\sigma}{dtdw^2}=\frac
{a(t,w^2)}{32~\pi^2},~~ s\rightarrow \infty,
\end{equation}
i.e. the differential cross section becomes $s$-independent, as it is expected 
for a $t-$channel exchange of a spin-one meson. In case of 
$\eta-$ or $\sigma-$exchange the cross section has to decrease with $s$ 
according to:
\begin{equation}
\frac {d^2\sigma}{dtdw^2}=\frac {f(t,w^2)}{s^2}.
\end{equation}
These different properties should help to sign experimentally the 
$\omega-$exchange contribution.

\subsection{The tensor analyzing power }

The predictions for this observable, in case of $\omega-$, $\eta-$ or 
$\sigma-$exchange  are very different and are given below. 

\underline{$\sigma$-exchange}

The $dd\sigma$-vertex can be described, in the general case, in terms of two 
independent form factors with the following spin structures: $ g_0(t)\vec 
U_1\cdot 
\vec U_2^*$ and $ g_2(t)\hat{\vec k}\cdot\vec U_1
~\hat{\vec k}\cdot \vec U_2^*.$, 
where $\hat{\vec k}$ is the unit vector along the three-momentum transfer, $\vec 
{U_1} (\vec U_2)$ is the vector 
of polarization of the initial (final) deuteron and $ g_0(t)$ and $ g_2(t)$ are 
form factors related to the deuteron form factors.

The tensor analyzing power is:
\begin{equation}
T_{20}=-\sqrt{2}g_2(t)\frac{2g_0(t)+g_2(t)}{3g_0^2(t)+g_2^2(t)+2g_0(t)g_2(t)}.
\end{equation}
It is possible to 
induce tensor analyzing power even in the case of $\sigma-$exchange (spin 0 
particle), only if one takes into account high order effects (interaction with 
derivatives) which are characterized by $g_2(t)$.

\vspace{.2 true cm}
\underline{$\eta$-exchange}

In this case the $dd\eta$-vertex is characterized by a single spin structure, 
$g_1(t)\hat{\vec k}\cdot\vec U_1\times\vec U^*_2$, and using the 
$NN^*\eta$-vertex in the form: $\chi_2^\dagger\vec \sigma\cdot \hat{\vec 
k}\chi_1$, one can find $T_{20}=\displaystyle \frac{1}{\sqrt{2}}$,  
positive and $t-$independent. The form factor $g_1(t)$ of the $dd\eta-$vertex 
is proportional to the magnetic form factor of the deuteron (in the impulse 
approximation).

\underline{$\omega$-exchange}

The tensor analyzing power in  $d+p\rightarrow d+X$, $T_{20}$, can be written 
in terms of the deuteron electromagnetic form factors as:
\begin{equation}
T_{20}=-\sqrt{2}\frac{V_1^2+(2V_0V_2+V_2^2)r(t)}
{4V_1^2+(3V_0^2+V_2^2+2V_0V_2)
r(t)},\label{li1}
\end{equation}
where  $V_0(t)$, $V_1(t)$ and $V_2(t)$ are linear combinations of the standard 
electric, $G_c$, magnetic $G_m$ and quadrupole $G_q$ deuteron 
form factors.
The ratio $r$ characterizes the relative role of longitudinal and transversal 
isoscalar 
excitations in the transition $\omega+N\rightarrow X$. It can be written, for 
the excitation of any nucleon resonance $N^*$,
as follows:
\begin{equation}
r(t)=\frac{|A_S^p+A_S^n|^2}{|A_{1/2}^p+A_{1/2}^n|^2+|A_{3/2}^p+A_{3/2}^n
|^
2}\equiv 
\sigma_L(t)/\sigma_T(t) ,
\end{equation}
where $A_S^{p,n}$ is the longitudinal form factor, $A_{1/2}^{p,n}$ and 
$A_{3/2}^{p,n}$ are the two 
possible transversal form factors, 
corresponding to 
total $\gamma^*+N$-helicity equal to $1/2$ and $3/2$, for proton and neutron.

The excitation of overlapping resonances with finite 
values of widths, introduces a $w$-dependence in $r$ as follows:
\begin{equation}
r(t)\rightarrow 
r(t,w)=\frac{\sum_i\sigma_{L,i}(t)B_i(w)C_i}{\sum_i\sigma_{T,i}(t)B_i(w)C_i},
\end{equation}
where $B_i(w)$ is a Breit-Wigner function for the $i-th~N^*$-resonance with a 
definite 
normalization:
\begin{equation}
C_i^{-1}=\int_{m+m_\pi}^\infty dw B_i(w),
\end{equation}
where $m_\pi$ is the pion mass.

The interference due to the excitation of different resonances does not appear 
in the inclusive processes. It is contained in the angular distribution of the 
decay products (for example $\pi$-production), on which an integration is 
implicitely 
performed in the previous formulas.

From Eq. (5) one can see that all information about the $\omega N N^*$-vertex 
is 
contained in the function $r$ only.
A zero value of $r$ results in a $t-$ and $w$-independent value for 
$T_{20}$, namely $T_{20}=-1/\sqrt{8}$, for any value of the deuteron 
electromagnetic form factors. The ratio $r$ is calculated using the 
collective string model in \cite{Bi94,Bi96,Bi97},
assuming $SU_{sf}(6)$ symmetry, including the contributions of the following 
resonances: $ N_{11}(1440),~S_{11}(1535),D_{13}(1520)~\mbox{ and} 
~S_{11}(1650)$,
which are overlapping in this energy region.

Of these four resonances only the Roper resonance has a nonzero isoscalar 
longitudinal electromagnetic form 
factor. The \underline{isoscalar} longitudinal amplitudes of $S_{11}(1535)$ 
and 
$D_{13}(1520)$ excitations vanish because of spin-flavor symmetry, while both 
isoscalar 
and 
isovector 
longitudinal couplings of 
$S_{11}(1650), D_{15}(1675)$ and $D_{13}(1700)$ excitations vanish identically.

In Fig. 2 we report the theoretical predictions for $T_{20}$, using Eqs. (5-6), 
together 
with 
the existing 
experimental data. In such approximation $T_{20}$ is a universal function of 
$t$ 
only, without 
any dependence on the initial deuteron momentum. The experimental values of 
$T_{20}$ for 
$p(\vec d,d)X$ \cite{lns250,Az96}, for different momenta of the incident 
beam 
are shown as 
open symbols. These data show a scaling as a function of $t$, with a small 
dependence on the 
incident momentum, in the interval 3.7-9 GeV/c. On the same plot the data for 
the elastic 
scattering process $e^-+d\rightarrow e^-+d$ \cite{t20} are shown (filled 
stars).

These different data show a very similar behavior: negative values, with a 
minimum in the region $|t|\simeq  0.35~GeV^2$ and their value increase toward 
zero at larger  $|t|$. The lines are the 
result of the $\omega$-exchange model for the $d+p\rightarrow d+X$ process: for 
$r=0$ (dashed-dotted line), the calculation based on \cite{Bi94} for the  Roper  
excitation only is represented by 
the dotted 
line and for the 
excitation of all the four resonances by the the full line.
The deuteron electromagnetic form factors have been taken from \cite{chu}, a 
calculation based 
on relativistic impulse 
approximation, and they reproduce well the $T_{20}-$data for $ed$ elastic 
scattering 
\cite{t20}.
When $r\gg 0$ or if the contribution of the deuteron magnetic form factor 
$V_1(t)$ is 
neglected, then $T_{20}$ does not depend on the ratio $r$, and coincides with 
$t_{20}$ for 
the elastic $ed$-scattering (with the same approximation).

From Fig. 2 it appears that the $t-$behavior of $T_{20}$ is very sensitive to 
the value of 
$r$ especially at relatively small $r$, $r\leq 0.5$. The values of $r$, 
predicted by 
model 
\cite{Bi94}, give a very good description of the data, when taking into 
account 
the 
contribution of all four resonances. These data, in any case, exclude a 
very 
small value of 
$r,~r \ll 0.1$ as well as very large values of $r$. Such sensitivity of 
$T_{20}$ for 
$d+p\rightarrow d+X$ to the ratio of the corresponding isoscalar form factors 
of 
the 
$N^*$-excitation gives an evident indication of the excitation of the Roper 
resonance in this process. The predicted dependence  of the 
ratio $r(t,w)$ and of $T_{20}$, on the initial deuteron momentum
is not so large, in agreement with experimental data. 

This model for $d+p\rightarrow d+X$ may be improved, taking into 
account for example, other 
meson exchanges, 
or the effects of the strong interaction in initial and final states. 
However the
corrections that can be added to the model presented here are strongly 
model and parameter dependent, and are not justified by the existing 
experimental data.

Let us note in this connection, that, in the considered model, all T-even 
polarization observables are 
nonzero and 
large in absolute value. This is an intrinsic property of $\omega$-exchange. 
But 
all T-odd 
polarization effects cancel, because we neglected the strong 
interaction in 
initial and final states. In case of collinear kinematics, all one spin T-odd 
polarization 
observables vanish, in any model. The most simple  T-odd polarization 
observable, which 
exists in the general case for the collinear kinematics, corresponds to the 
following 
correlation coefficient: $\hat{\vec k}\cdot\vec P\times\vec Q$, $Q_a=Q_{ab}k_b$, 
where $\vec P$ is the proton 
polarization and $Q_{ab}$ 
is the deuteron tensor polarization. A measurement of these observables will 
give a direct 
information on the presence and intensity of the final or initial strong 
interaction. The 
tensor analyzing power $T_{20}$, being a T-even observable, is less sensitive 
to such effects.

\subsection{ Coefficients of polarization transfer in {\boldmath $\vec 
d+p\rightarrow \vec d+X$}}

Let us calculate the (vector) transfer polarization coefficients $K_a^{a'}$ 
(with $a$, $a'=x,y$ or $z$) from initial to final deuterons, in the processes   
$\vec 
d+p\rightarrow \vec d+X$, in the framework of  $\omega-$ 
exchange. Using the corresponding 
parametrization of the $\omega dd-$vertex, one can find the following 
formulas, 
for the nonzero polarization transfer coefficients:
\begin{eqnarray}
&K_y^{y'}=K_x^{x'}=\displaystyle\frac {3}{2} 
\frac{V_1^2+(V_0V_2+V_0^2)r(t)}{4V_1^2+(3V_0^2+V_2^2+2V_0V_2)r(t)},\\
&K_z^{z'}=\displaystyle \frac {3}{2} 
\frac{V_1^2+V_0^2r(t)}{4V_1^2+(3V_0^2+V_2^2+2V_0V_2)r(t)}.
\end{eqnarray}
For $r=0$ we obtain simply: $K_x^{x'}=K_y^{y'}=K_z^{z'}=\frac {3}{8}=0.375$ : 
all coefficients are positive and $t-$independent. As in the case of $T_{20}$ 
the ratio $r(t)$ contains all the information about the properties of $\omega 
p\rightarrow X$ vertexes.
From Fig. (3), one can see that $K_y^{y'}$ has a strong $q$ and $r$ 
dependence. 
The point where all lines for different $r$ are crossing is due (as for 
$T_{20}$) to a special combination of $V_0$ and $V_2$ (see the numerator of 
Eq. (6)). At $q=0$, $K_y^{y'}=1/2$, for any value of $r$, as $V_1=V_2=0$. The 
largest sensitivity to $r$ is in the region $q\geq 3~fm^{-1}$ and the position 
of zero crossing is strongly $r-$dependent.  

\section{Polarization phenomena in 
{\boldmath 
$\lowercase{\vec p}+\lowercase{d}\rightarrow  
\lowercase{\vec p}+M^0+\lowercase{d},~M^0=\sigma$} or \boldmath $\pi$ }

The polarization properties of the produced protons in the processes $\vec 
p+d\rightarrow \vec p + M^0 +d, ~M^0=\sigma,~\eta$ or $\pi^0$ depend 
essentially 
on the kind of the produced meson and on the quantum numbers (${\cal J}^P $) 
of 
the nucleonic resonance in the intermediate state : $p+d\rightarrow N^*
({\cal J}^P )+d\rightarrow p + M^0 +d$.
Let us consider the case of Roper excitation, with ${\cal 
J}^P={\frac{1}{2}}^+$. 
\subsection{$\omega$-exchange}

The nonzero polarization transfer coefficients for 
$\sigma$-production: 
$\vec 
p+d\rightarrow \vec p +\sigma +d$ are :
\begin{equation}
\displaystyle K_y^{y'}=K_x^{x'}=\frac {\cal R}{4+\cal R}, ~~K_z^{z'}=-\frac 
{-4+ 
\cal R}{4+\cal R},
~{\cal R}^{-1}=\frac{V^2_1}{(3V_0^2+V_2^2+2V_0V_2)r(t)},\label{li1}
\end{equation}
i.e. these coefficients depend only on the momentum transfer $t$.

On the other hand a dependence from the $M^0$-production angle is contained in 
the decays $N^*\left( {\frac{1}{2}}^+\right )\rightarrow p+\pi^0$ (or 
$p+\eta$), 
with $P$-wave production of the pseudoscalar meson. In this case we obtain 
that 
all $K_a^{a'}$ coefficients allowed by the P-invariance of strong interaction 
are nonzero:
\begin{eqnarray}
&K_y^{y'}=-\displaystyle \frac {\cal R}{4+\cal R},\\
&K_x^{x'}=(1-2cos^2\theta_{\pi})\displaystyle\frac {\cal R}{4+\cal R},\\
&K_z^{z'}=(1-2cos^2\theta_{\pi})\displaystyle\frac {4- \cal R}{4+\cal R},\\
&K_x^{z'}=(sin~2\theta_{\pi})\displaystyle\frac {-4+ \cal R}{4+\cal R},\\
&K_z^{x'}=(sin~2\theta_{\pi})\displaystyle\frac {\cal R}{4+\cal R},
\end{eqnarray}
where we used a coordinate system with the $z-$axis along the momentum 
transfer 
$\vec k$, $y\|\vec n$, with $\hat{\vec n} =
\vec k\times \vec q / |\vec k\times\vec q |$, ($\vec q$ is the 
meson 3-momentum) and $x \| \vec n\times \vec k $.

An important property of the $\omega -$exchange model is the universal 
dependence of all the  $K_a^{a'}$ from $t$, through the ratio ${\cal R}(t)$.
An experimental check of the $\theta_{\pi}$-dependence of the polarization 
transfer coefficient would be a signature of the validity of this model.

\subsection{$\sigma$-exchange}

In the case of $\sigma-$exchange it is possible to analyze the spin transfer 
coefficients in a general form, with a model independent parametrization, for 
two possible generalized vertexes: $\sigma +N\rightarrow \sigma +N$ and  $\sigma 
+N\rightarrow \pi^0 +N$.
The spin structure of the corresponding matrix elements can be written as 
follows:
\begin{eqnarray*}
&F(\sigma N \rightarrow \sigma N) =\chi^{\dagger}_2(A+i\vec\sigma\cdot\hat{\vec 
n} 
B) 
\chi_1,\\
&F(\sigma N \rightarrow \pi N) =\chi^{\dagger}_2(C\vec\sigma\cdot\hat{\vec k}+ 
D\vec\sigma\cdot\hat{\vec q}) \chi_1,
\end{eqnarray*}
where $A,~ B,~ C,$ and $D$ are the corresponding amplitudes, depending from 
three independent kinematical variables: $w$ (the invariant effective mass of 
the produced $MN-$ system), $cos \theta_{\pi}=\hat{\vec k}\cdot\hat{\vec q}$ and 
$t$. 
In case of $\sigma-$ exchange the spin transfer coefficients depend only from 
the 
ratio 
$r_g(t)=g_2(t)/g_0(t)$ of the form factors of the $dd\sigma$-vertex. For the 
nonzero (diagonal) coefficients the following expressions hold:
\begin{equation}
K_y^{y'}=K_x^{x'}=\displaystyle\frac{3(1+r_g)}{2(3+2r_g+r^2_g)},~K_z^{z'}=
\displaystyle\frac{3}{2(3+2r_g+r^2_g)},
\label{li1}
\end{equation}
where the $z-$axis is along the 3-momentum transfer $\vec k$. In this case the 
observables $K_a^{a'}$ contain the same information as $T_{20}$, Eq. (5). 
Moreover these observables are connected by the following relation:
\begin{equation}
T_{20}-\frac{3}{\sqrt{2}}K_z^{z'}+\frac{9}{2\sqrt{2}}(K_y^{y'})^2=0.
\label{li1}
\end{equation}

The polarization properties 
of 
both protons, in the initial and final states, do not depend from the 
structure 
of the $dd\sigma$-vertex, but depend on the amplitudes $A$ and $B$ for 
$\sigma$- 
production (or on the amplitudes $C$ and $D$ for $\pi$ production).

The vector polarization $\vec P$ of the protons, produced in the collision of 
unpolarized particles has to be different from zero, in the general case:
$$\vec P=\hat{\vec n} \displaystyle\frac {2{\cal I}m 
~AB^*}{|A|^2+sin^2\theta_{\pi} 
|B|^2}$$

\noindent for $\sigma-$production, and 
$$\vec P=\hat{\vec n} \displaystyle\frac {2{\cal I}m~ 
CD^*}{|C|^2+2cos\theta_{\pi}{\cal R}e~ CD^*+|D|^2}$$
 
\noindent in case of $\pi(\eta)$-production.
The analyzing powers in $\vec p+d \rightarrow p+M+d$ are determined by the 
evident relation: ${\cal A}_y=P$.

The dependence of the polarization $\vec P_2$ of the produced proton on the 
polarization $\vec P_1$ of the proton beam is described by the following 
general 
formulas:
$$ \vec P_2=\vec P_1 
\displaystyle\frac{|A|^2-sin^2\theta_{\pi}|B|^2}
{|A|^2+sin^2\theta_{\pi}|B|^2}
+2\hat{\vec n}(\hat{\vec n}\cdot \vec P_1)
\displaystyle\frac{ |B|^2}
{|A|^2+sin^2\theta_{\pi}|B|^2} 
-2(\hat{\vec n}\times \vec P_1)
\displaystyle\frac{ {\cal R}e  AB^*}
{|A|^2+sin^2\theta_{\pi}|B|^2},$$
\noindent for  $\sigma$-production, and
$$\vec P_2=-\vec P_1 +
\displaystyle\frac{2\hat{\vec k} (\hat{\vec k}\cdot \vec P_1)|C|^2}
{|C|^2+2 {\cal R}e C D^*cos\theta_{\pi}+|D|^2}
+2\hat{\vec k}(\hat{\vec k}\cdot \vec P_1)
\displaystyle\frac{ |D|^2}
 {|C|^2+2 {\cal R}e~ C D^*cos\theta_{\pi}+|D|^2}$$
\noindent
 for $ \pi^0(\eta)$-production.

From these formulas one derives the following expressions for the coefficients 
$K_a^{a'}$:

\noindent 1. \underline{$\sigma$-exchange with $\sigma$-production} ($\vec 
p+d\rightarrow \vec 
p+\sigma 
+d$):
\begin{eqnarray*}
\displaystyle 
&K_y^{y'}=+1,\\
&K_x^{x'}=K_z^{z'}=\displaystyle\frac{|A|^2-sin^2\theta_{\pi}|B|^2}
{|A|^2+sin^2\theta_{\pi}|B|^2},\\
&K_x^{z'}=K_z^{x'}=2sin^2\theta_{\pi}\displaystyle\frac{ {\cal R}e~ AB^*}
{|A|^2+sin^2\theta_{\pi}|B|^2}.
\end{eqnarray*}

\noindent 2. \underline{$\sigma$-exchange with $\pi^0$-production} 
($p+d\rightarrow p+\pi^0(\eta) 
+d$):
\begin{eqnarray*}
\displaystyle 
&K_y^{y'}=-1,\\
&K_x^{x'}= -1 +\displaystyle\frac{2sin^2\theta_{\pi}|D|^2} {|C|^2+2 {\cal R}e~ 
C 
D^*cos\theta_{\pi}+|D|^2},\\
&K_x^{z'}=K_z^{x'}=\displaystyle\frac{ sin^2\theta_{\pi}|D|^2}{ |C|^2+2 {\cal 
R}e ~C D^*cos\theta_{\pi}+|D|^2}.
\end{eqnarray*}
\subsection{Case of interfering resonances}
The formulas, previously derived, are general and they can be applied to any 
resonance.
The amplitudes $A,~B,~C$ and $D$ depend on the quantum numbers of the ${\cal 
J}^P$ of the excited nucleonic resonances.
Let us consider two cases:
\begin{itemize}
\item[$(a)$]  RR excitation, with ${\cal J}^P= \displaystyle\frac{1}{2}^+$, 
\item[$(b)$] $N^*(1535)$ excitation, with  ${\cal J}^P= 
\displaystyle\frac{1}{2}^-$.
\end{itemize}

\noindent 1. \underline{$\sigma$-exchange with with $\sigma$-production} 
($p+d\rightarrow p+\sigma +d$):

We obtain the following amplitudes:

$(a)$ : $A=f^{(+)}(t){\cal H}^{(+)}(w)g^{(+)}_{\sigma},~~~ B=0$.

$(b)$ : $A= cos\theta_{\pi}B,~~~ B=f^{(-)}(t){\cal 
H}^{(-)}(w)g^{(-)}_{\sigma}$,

\noindent where $g^{(\pm)}_{\sigma}$ is the constant for the decay $N^*\left 
({\cal J}^P= \displaystyle\frac{1}{2}^{\pm}\right )\rightarrow N+\sigma $, 
${\cal H}^{(\pm)}(w)$ is the corresponding propagator to decribe the Breit 
Wigner behavior:
$${\cal H}^{(\pm)}(w)=\displaystyle \frac{1}{w-m^*-i\frac{\Gamma}{2}}$$
for the excitation of $N^*$ with mass $m^*$ and total width $\Gamma$; 
$f^{(\pm)}$ are form factors of the vertexes $p\rightarrow N^*+\sigma $.

\noindent 2. \underline{$\sigma$-exchange with $\pi^0$-production} 
($p+d\rightarrow p+\pi^0(\eta) 
+d$):

$(a)$ : $C=0$, $D=f^{(+)}(t){\cal H}^{(+)}(w)g^{(+)}_{\pi}$. 

$(b)$ : $C=f^{(-)}(t){\cal H}^{(-)}(w)g^{(-)}_{\pi}$,~$D=0$,

\noindent where $g^{(\pm)}_{\pi}$ is the constant for the decay $N^*\left 
({\cal 
J}^P= \frac{1}{2}^{\pm}\right )\rightarrow N+\pi$.

We can conclude that any single resonance induces zero analyzing power in the 
processes $\vec p+d \rightarrow p+\sigma(\pi)+d$. Nonzero analyzing powers 
result from the interference of different resonances contributions. On the 
other 
hand the vector polarization transfer coefficients are nonzero for a single 
resonance and depend strongly on ${\cal J}^P$ and on the produced meson.
Using the general formulas obtained above, we find for the polarizations:

\noindent 1. \underline{$\sigma$-exchange with $\sigma$-production} ($\vec 
p+d\rightarrow \vec 
p+\sigma 
+d$):

$(a)$ : $\vec P_2=\vec P_1$.

$(b)$ : $\vec P_2= (-1+2cos^2\theta_{\pi})\vec P_1+2\hat{\vec n}(\hat{\vec n} 
\cdot \vec 
P_1)+ 2cos\theta_{\pi}\vec P_1\times\hat{\vec n}$. 

\noindent 2. \underline{$\sigma$-exchange with $\pi^0$-production} ($\vec 
p+d\rightarrow \vec 
p+\pi^0(\eta)  +d$):

$(a)$ : $\vec P_2=-\vec P_1+ 2\hat{\vec q}(\hat{\vec q}\cdot \vec P_1)$.

$(b)$ : $\vec P_2= -\vec P_1+ 2\hat{\vec k}(\hat{\vec k}\cdot \vec P_1)$.
\subsection{Polarization effects in case of interfering resonances}
In conclusion of this section we give the results for the polarization 
properties of protons in processes $p+d\rightarrow p+M^0 +d$, 
induced 
by the {\bf interference of two nucleonic resonances} with ${\cal J}^P= 
\frac{1}{2}^+$, 
and ${\cal J}^P= \frac{1}{2}^-$, in the framework of the $\sigma-$exchange 
model.

\noindent 1. \underline{$\sigma$-production} ($\vec p+d\rightarrow \vec 
p+\sigma 
+d$):

$(a)$ : $\vec P_2=\hat{\vec n} {\cal A}_{\sigma},~  
\displaystyle\frac{d\sigma}{d\omega} =\left (\frac{d\sigma}{d\omega} \right 
)_0(1+{\cal A}_{\sigma}\hat{\vec n}\cdot \vec P_1),~{\cal A}_{\sigma}\simeq 
2{\cal 
I}m 
~cd^*,$

$(b)$ : $\vec P_2\simeq 2\hat{\vec n} \times \vec P_1{\cal R}e~cd^*,$
where:
$$c=f^{(-)}(t){\cal H}^{(-)}(w)g^{(-)}_{\sigma},$$
$$d=f^{(+)}(t){\cal H}^{(+)}(w)g^{(+)}_{\sigma}+cos\theta_{\pi}c.$$

\noindent 2. \underline{$\pi^0 (\eta)$-production} ($\vec p+d\rightarrow \vec 
p+\pi^0(\eta)  +d$):

$(a)$ : $\vec P_2=\hat{\vec n} {\cal 
A}_{\pi},~\displaystyle\frac{d\sigma}{d\omega} 
= 
\left (\frac{d\sigma}{d\omega}\right )_0(1+{\cal A}_{\pi}\hat{\vec n}\cdot \vec 
P_1),~{\cal A}_{\pi}\simeq 2{\cal I}m~ ab^*,$

$(b)$  : $\vec P_2=\simeq 2\left [-cos\theta_{\pi}\vec P_1+\hat{\vec q}\hat{\vec 
k}\cdot 
\vec P_1+\hat{\vec k}\hat{\vec q}\cdot \vec P_1\right ]{\cal R}e~ab^*,$

\noindent where we used the following notations:
$$a=f^{(+)}(t){\cal H}^{(+)}(w)g^{(+)}_{\pi},$$
$$b=f^{(-)}(t){\cal H}^{(-)}(w)g^{(-)}_{\pi}.$$
These formulas show that the T-odd analyzing power in $\vec p+d$-collisions (or 
the polarization of the final proton, produced in the collision of unpolarized 
particles) are especially sensitive to the interference of different nucleonic 
resonances.
 
\section{Remarks on polarization observables in the process {\boldmath \reac}}

The reaction \reac ~ has been considered a good tool for the study of the Roper 
resonance \cite{Mo92} as the $\alpha$ particle, is known to be a good isoscalar 
probe. Calculations on this reaction have been done in \cite{Hi96} and 
polarization observables analyzed in \cite{Hi99}, in framework of 
$\sigma$-exchange.

Two main features characterize this reaction: it is a three body reaction, in 
the final state, with non coplanar kinematics and only protons have nonzero 
spin.
The acoplanarity is related to the following combination of 3-momenta : 
$a=\displaystyle\frac{\vec 
q\cdot\vec p_1\times 
\vec p_2}{E_1E_2E_{\pi}}$, where $\vec p_1$ and  $\vec p_2$ are the three 
momenta of the initial 
and final proton, $\vec q$ is the 3-momentum of the produced pion and $E_1$, 
$E_2$, $E_{\pi}$ the corresponding energies. This 
combination is present in the full set of the 5 independent kinematical 
variables which are necessary for the complete description of a 
process $1+2\to 3+4+5$.

The variable $a$ is connected with the azymuthal angle $\phi$- between two 
different reaction planes: 
the scattering plane of the proton (i.e. the plane defined 
by the 3-momenta $\vec p_1$ and $\vec p_2$ ) and  the plane defined 
by the pion three-momentum $\vec q$ and the transferred momentum $\vec p =\vec 
p_1 - \vec p_2$. 
The polarization of the final proton can be parametrized in the 
following general form:
\begin{equation}
\vec P=\hat{\vec n} P_n +a(\hat{\vec m} P_m+\hat{\vec k} P_k),
\end{equation}
where the unit vectors $\hat{\vec m}$, $\hat{\vec n}$ and $\hat{\vec k}$ are 
redefined in this section as:
$\hat{\vec n}=\vec p_1\times \vec p_2/|\vec {p_1}\times\vec {p_2}|$, 
$\hat{\vec k}=\vec p_1 / |\vec p_1|$,
$\hat{\vec m}=\hat{\vec n}\times \hat{\vec k}$. $P_n$,  $P_m$ and $P_k$ are 
the three 
independent 
and nonzero components of the final proton polarization vectors.
If the P-invariance of the strong interaction holds, the matrix element is 
described by the following general parametrization (in the CMS of the considered 
reaction):
\begin{equation}
{\cal M}=\chi^\dagger_2\left [ \vec \sigma\cdot\vec m 
f_1+\vec\sigma\cdot\hat{\vec k} 
f_2+a\left (i\tilde f_1+\vec\sigma\cdot\hat{\vec n} \tilde f_2\right )\right 
]\chi_1, 
\end{equation}
where $\chi_1$ and $\chi_2$ are the 2-component spinors of the 
protons in the 
initial and final states; $f_1$, $ f_2$, $\tilde f_1$ and $\tilde f_2$ are the 
scalar independent amplitudes for \reac, which are functions of the 5  
kinematical variables, defined above.

Nine 
coefficients of polarization transfer, $D_{ij}$ can be defined in terms of the 
scalar amplitudes $f_i$ and 
$\tilde f_i$. Let's focus on $D_{nn}$.
$$D_{nn}\left (\displaystyle\frac{d\sigma}{d\omega}\right )_0
=-|f_1|^2-|f_2|^2+a^2\left (|\tilde f_1|^2+|\tilde f_2|^2\right ).$$

In collinear kinematics only a single and specific spin structure is present: 
all polarization phenomena can be predicted exactly, in a model 
independent form:
$$D_{mm}=D_{nn}=-D_{kk}=-1,$$
and all other polarization observables have to be identically zero, due to the 
azimuthal symmetry of the collinear kinematics.

In the general case, two main mechanism contribute to the matrix element for 
\reac , one related to the projectile and the other to the target excitation. 
Following \cite{Hi99}, the Deck mechanism results from $\pi$-exchange, but the 
Roper excitation is induced by $\sigma$-exchange. In this hypothesis, one can 
show that both non-coplanar amplitudes $\tilde f_1$ and $\tilde f_2$ are zero 
for both mechanisms, taking into account the most general properties of 
$\pi\alpha$-scattering (for the Deck mechanism) and of the process 
$\sigma+N\to N+\pi$ (for the Roper excitation). From the spin structure 
given above, one can deduce that $D_{nn}=1$, in the whole region of kinematical 
variables (for coplanar and non-coplanar kinematics). The polarization of the 
final proton has only one non-zero component, in the 
$\hat{\vec n}$-direction, i.e. along the normal to the proton scattering plane. 
and 
its sign and absolute value  depend on the relative role 
of 
the considered mechanisms i.e. to the details of 
the corresponding amplitudes. 

This 'coplanar-like' behavior of $\sigma-$ and $\pi-$ exchanges in 
\reac  ~is related to the fact that these mediators are 
spinless particles. Such exchanges can not connect different reaction planes. 
This 
conclusion does not depend on details, approximations, values of the 
constants or shape of form factors which are typically taken in the numerical 
applications, because it is based only on the value of the spin of the 
mediators.

Earlier we suggested that the $\omega-$ exchange is the most probable mechanism 
for the Roper excitation, in this energy range \cite{Re96}. The most important 
difference with respect to $\sigma$-exchange is due to the spin and 
has evident implications for the polarization phenomena: a vector particle 
exchange induces all four amplitudes different from zero, in the general case.

Future experimental data on polarization observables for \reac ~will constitute 
a 
crucial test in order to disentangle the mechanisms involved.
\section{Conclusion}

Let us summarize the main results of the previous analysis of polarization
phenomena for the Roper excitation in $dp-$ and $\alpha p$-collisions, 
in the framework of  $\omega-$ exchange,   that we consider as the most probable 
mechanism for these processes.
\begin{itemize}
\item Polarization effects are rich in information, especially for 
correlation experiments,
like $d+p\rightarrow d+p+M$, where M is $\pi-$ , $\eta$- or $\sigma-$meson.

\item The $t-$dependence of the existing data on tensor analyzing power in 
$d+p\to d+X$ is sensitive to the value of the spin of exchanged particles and 
agrees very well with the model based on $\omega-$ exchange.

\item The contributions of the four overlapping resonances, $N_{11}(1440)$, 
$S_{11}(1535)$, 
$D_{13}(1520)$ and $S_{11}(1650)$, have to be taken into account, to reproduce 
the 
$t-$dependence of the tensor analysing power in $d+p \to d+X$.
The longitudinal form factor of the Roper excitation is essential for a correct 
description of experimental data.

\item Polarization phenomena in $p+d \to p+d+M$  are sensitive to the type of 
the produced meson M and to the quantum numbers of mediators.

\item The acoplanarity of processes with three particles in the final state,  
introduces many new interesting aspects in the
analysis of the spin structure of matrix elements and polarization phenomena.

\item  The spin structure of matrix element for $p+\alpha \to p+\pi+\alpha$ is 
characterized by four scalar
amplitudes -with two non-coplanar amplitudes, so that the polarization vector of
scattered protons has, in general, all three nonzero components.

\end{itemize}

\begin{figure}
\caption{ Possible mechanisms for $d+p\rightarrow d+\pi+N$:  
(a) deuteron excitation;  (b) proton excitation.}
\end{figure}
\begin{figure}
\caption{Experimental data on $T_{20}$ for $ d+p\rightarrow d +X$ 
at incident momenta of 
3.75 GeV/c  (open diamond) \protect\cite{lns250}
5.5 GeV/c  (open circles), 
4.5 GeV/c  (open squares), and  9 GeV/c (open triangles)
\protect\cite{Az96}. 
The prediction of the $\omega-$exchange model for  $r=0$ is represented by the  
dashed-dotted line.
The calculation with $r$ from \protect\cite{Bi94} is represented by the 
dotted 
line for the  Roper  excitation 
and by the solid line for the excitation of the four four resonances mentioned 
in the text. The $t_{20}$ data 
from $ed$ elastic scattering (filled stars)
are from \protect\cite{t20}.}
\end{figure}

\begin{figure}
\caption{Vector polarization transfer coefficient $K_y^{y'}$ as a 
function of 
$q$, for $ d+p\rightarrow d +X$. The calculation with $r$ from 
\protect\cite{Bi94} is represented by the 
dotted 
line for the Roper excitation 
and by the solid line for the excitation of the four resonances mentioned in the 
text.}
\end{figure}

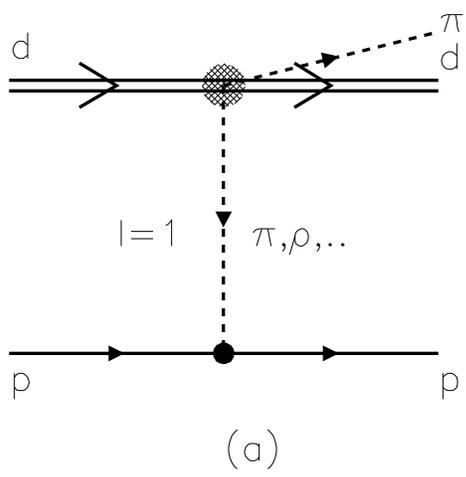
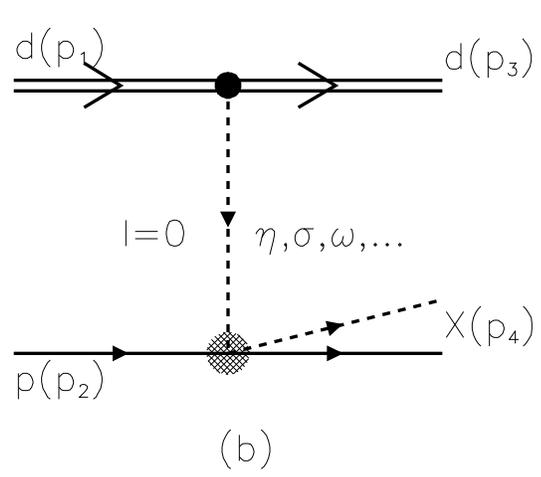
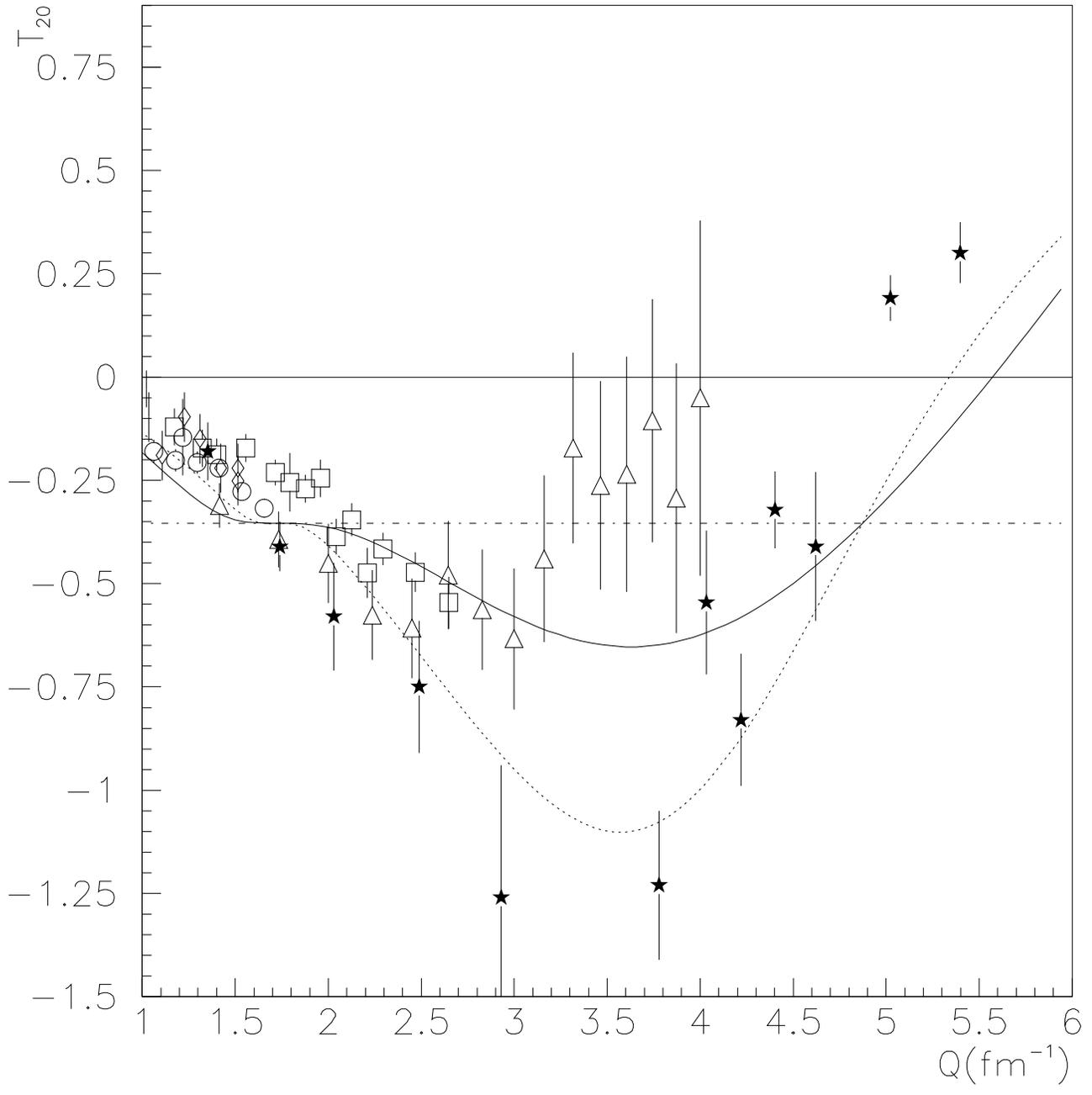
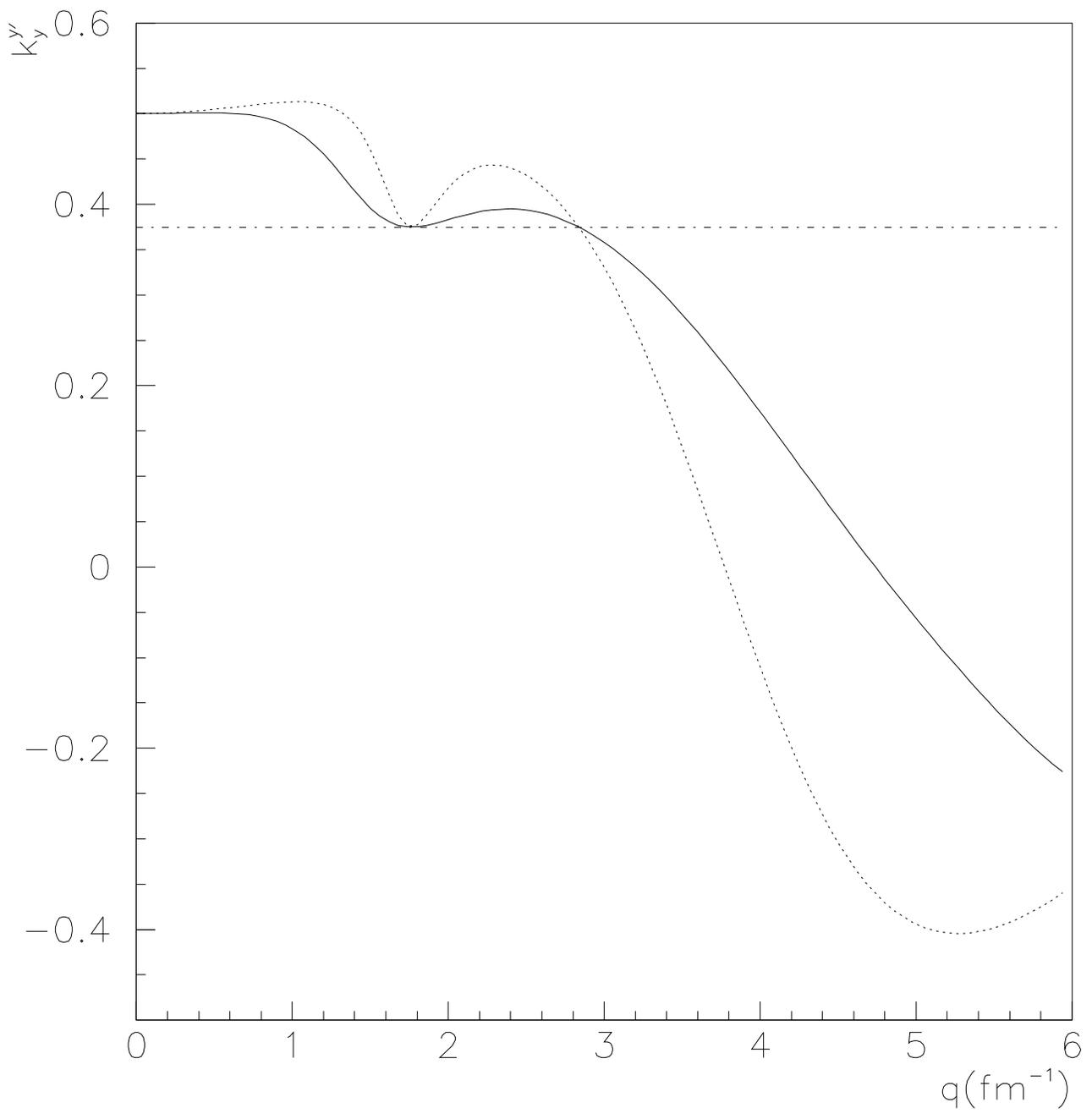
\end{document}